# Strong Adsorption between Uranium Dicarbide and Graphene Surface Induced by f Electrons

Jie Han,[†] Xing Dai,[†] Cheng Cheng,[‡] Minsi Xin,[†,§] Zhigang Wang,*,[†,§] Ping Huai,*,[‡] and Ruiqin Zhang*,[§,#]

[†]Institute of Atomic and Molecular Physics, Jilin University, Changchun 130012, P. R. China

[‡]Shanghai Institute of Applied Physics, Chinese Academy of Sciences, Shanghai 201800, P. R. China

[§]Beijing Computational Science Research Center, Beijing, P.R. China

[#]Department of Physics and Materials Science and Centre for Functional Photonics (CFP), City University of Hong Kong, Hong Kong SAR, P. R. China

The interaction between gaseous uranium dicarbide and graphite is significant for the safety control and design of Gen-IV nuclear energy system. In this article, the interaction mechanism has been studied using a simplified model of adsorption of two typical $UC_2$ isomers (linear CUC and symmetric triangular structures) on graphene based on density functional theory calculations. The results reveal strong chemisorption characteristics between the $UC_2$ and graphene, which is found different from the conventional weak intermolecular interaction. Interestingly, although the CUC structure can induce a double $sp^3$-hybridization at the graphene, the most stable adsorption structure is formed by the triangular $UC_2$ adsorbed at the hollow site of the graphene. Further bonding analysis indicates that the U 5f orbitals of the triangular $UC_2$ are more active than that in the CUC, providing a larger effective bonding area in the adsorption system. Our calculations are helpful for understanding the role of actinide compounds in adsorption on carbon nanomaterials surface, especially for elucidating the bonding properties of 5f electrons.

## ■ INTRODUCTION

Graphite materials[1] are popularly used as neutron moderator or reflector in various nuclear reactors due to their superior properties such as high moderation ratio, low coefficient thermal expansion and satisfactory radiation stability. In high temperature gas cooled reactors (HTGRs), the coated fuel particles are embedded in matrix graphite.[2-4] At present, in Canadian deuterium uranium reactor (CANDU) and Gen-IV HTGRs, graphite is an essential component of coated fuel particles, in which pyrolytic carbon (PyC) layers serve as buffer or barrier of gaseous fission product, and also protect the cladding (SiC or alloy) from possible mechanical interactions and chemical attack.[5-7]

The interaction between nuclear fuel materials (nuclear fuel or nuclear waste) and nuclear graphite materials is a crucial issue to deal with to guarantee no detrimental effect in nuclear system. This is because the damage if any would affect the reaction rate and limit the service time of the reactor, and even lead to radioactive leakage under a serious condition.[6-8] Related research work has been performed on actinide oxides, molten salt and other nuclear fuel materials.[7,9,10] Owing to their outstanding physical properties such as high burn up, high linear power and high thermal conductivity,[11-13] uranium carbides are considered to be one of the most ideal candidates of nuclear fuel materials in Gen-IV nuclear reactor and are mostly used in coated fuel particles.[14-16] Previous studies showed that the uranium carbides could exist in the forms of UC, $UC_2$ and $U_2C_3$,[17,18] and vaporization of uranium carbides could occur during their use as nuclear fuels,[19,20] in which the $UC_2$ is the most abundant gaseous molecule identified by subsequent experimental work.[21] Mass spectroscopy has also evidenced the presence of gas $UC_2$ over solid $UC_2$.[22] Therefore, the $UC_2$ molecules can directly contact with the graphite; however, the mechanism of interaction has not been explored. Due to the radioactivity of nuclear materials, the experiments for nuclear fuel can only be carried out by remote manipulation in well shielded hot cells.

Actually, a small fraction of coated fuel particles may be defective, facilitating direct release of fission products from the fuel particles to the matrix graphite,[6] and the quality and failure evaluations of the coated fuel particle depends on the structure and properties of pyrolytic graphite; however, due to the incomplete understanding on the variant conditions, researchers cannot get a proper prediction result.[5,23] Therefore, uncovering the interaction mechanism between graphite materials and uranium carbide molecules is significant for evaluating the quality of coated fuel particles and the safety of reactors.

Although the experimental research of interaction between nuclear fuels and graphite materials is constrained due to the radioactivity of nuclear materials, related research work has consistently been active. The researches conducted in the last century basically found no corrosion or detrimental damage between graphite and molten salt fuel (LiF-BeF$_2$-UF$_4$) or $UO_2$ fuel.[7,9] Recently, Cerefice et al. found that the adsorption of the depleted uranium ($UO_2(NO_3)_2$) onto graphite was not insignificant in the transport of uranium and the adsorption mechanism was unknown.[10] Carbon nanomaterials could form metallic functional materials with atoms or compounds containing f electrons (such as 4f element[24] and 5f element[25-31]), providing complex electronic structures and high chemical reactivity. Their interaction with the actinide compounds will certainly need further experimental and theoretical researches, for the consideration of nuclear and environment safety and other aspects.

It would be cost effective to conduct a theoretical study on the interaction characteristics between the uranium dicarbide and graphite before an expensive experiment. Because of the complexity and diversity of nuclear graphite materials, the adsorption energies obtained by density functional theory (DFT) based on a periodic multi-layer graphene are still not at a level of accuracy to enable quantitative modeling.[32] For the sake of generality and considering the capability of first-principles calculations, one would choose a graphene fragment as a simplified model of a single sheet nuclear graphite. Moreover, the perfect graphene with all $sp^2$-hybridization is more stable than other structures with defects and offers unique mechanical and chemical properties. Thus, the simplified model is suitable for a fundamental study of the interaction between uranium dicarbide molecules and graphite materials to investigate the qualitative adsorption trends and the mechanism. The $UC_2$ is a representative and most possible uranium carbide molecule adsorbed on the graphene in nuclear reactor. Though only linear CUC structure (triplet ground state) presents in experiments, the symmetric triangular $UC_2$ structure (quintet ground state) is the energetically most favorable in calculations by first principles theory.[20,33,34] On the other hard, since the highest occupied molecular orbitals are four single-electron orbitals with uranium 7s and 5f characters in the symmetric triangular $UC_2$,[20] and in uranofullerene system 7s electrons will transfer to carbon cage, 5f orbitals have contributions to the covalent interactions with the carbon cage.[28,30] A strong interaction between symmetric triangular $UC_2$ and graphene can be predicted with the participation of U atom. However, the frontier orbitals in the linear CUC are almost localized on U-C triple bonds,[33] its interaction with graphene needs further investigation. (See more information of the $UC_2$ reported in previous research [19,20,33,34] and obtained in our work in part 1 of the Supporting Information) Herein, we present a comprehensive study of the adsorption characteristics of these two representative $UC_2$ structures on graphene. Based on first-principles calculations using the DFT, we obtained the chemisorption feature from the analyses of electron population, bonding properties and also the vibration spectrum. Our calculations show apparent chemisorption characteristics in the system with the participation of 5f electrons, which provides an important reference for nuclear safety control and design.

■ **COMPUTATIONAL METHODOLOGY**

The computation of systems containing actinide elements requires the use of reliable methods which take the relativistic effect and electronic correlation into account. Qualified ab initio methods include a complete active space (CAS) SCF method which offers high accuracy but is computationally costly, making it only suitable for very small molecules.[35] A more suitable method is the density functional theory (DFT)[36] which can take the electron correlation into account with relatively small computational cost and has been widely used for studying complex systems. In the framework of DFT, relativistic effective core potential (RECP)[37-39] could describe the core electrons for high Z elements, which can further reduce the computational cost to some extent, ideal for calculating the models we constructed for the adsorption of $UC_2$ on the surface of graphene which contain a considerably large number of atoms (up to 123).

We adopted a hybrid PBE functional[40-42] which had been successfully used in the previous theoretical calculation of uranium carbide molecules.[20,33] We have done further test (See part 1 in the Supporting Information), and obtained results consistent with those in reference [20] based on the Complete Active Space with Second-order Perturbation Theory (CASPT2)[43,44], no matter using 60 or 78 core electrons for U (3-21G for C). Considering the capability to give results agreeing well with experiments, we chose to use a small-core relativistic effective core potential of the Cologne Stuttgart group (ECP60MWB)[45,46] in conjunction with a 10s9p5d4f3g contracted from 14s13p10d8f6g valence basis set (ECP60MWB_ SEG basis)[46] for U and a double zeta basis set (i.e. 3-21G)[47] for C and H, following an economic basis set scheme[48]. The approach has been demonstrated to be advantageous in determining the structure and charge population of similar systems,[27,49] and a larger basis set 6-31G has also been used for a further verification (see part 2 of the Supporting Information). In order to further ensure the reliability of the results, we constructed a graphene fragment (consisting of 120 atoms including 96 C with 24 H to saturate the fragment edge) which is large enough to ensure the behavior of its center similar to the bulk graphene.[50] The model was found especially reasonable in a research of strong local interaction[51]. Above the center, the $UC_2$ molecule was placed to simulate the adsorption. To closely relate with practically possible adsorption situations, we considered a variety of possible initial configurations with two typical isomers of $UC_2$ (i.e. a linear CUC and a symmetric triangular structures) respectively adsorbed on the graphene surface at a distance of about 3.5 Å. In the optimization, the systems were fully relaxed, followed by verifications using vibration frequency analysis at the same level of theory to ensure that the structures we obtained are at the real local minima on the potential energy surface and to estimate the zero-point energy (ZPE). For each system, electronic states with different spin multiplicities have been calculated. Since the results do not alter the conclusions and considering that the focuses here are on adsorption behavior and mechanism, we report here mainly the ground state of each adsorption structure and give the results of ZPE corrections and relative electronic states in Table S5 of the Supporting Information. All calculations were performed using Gaussian 09 package[52].

■ **RESULTS AND DISCUSSION**

After structure optimizations, we obtained four kinds of energetically favorable systems, in which the $UC_2$ adsorb on the graphene as shown in Figure 1. All these structures present a unique characteristic feature that the U is located at the hollow site of graphene. The linear CUC isomer is not stable on graphene, and the cluster converts to a bent CUC geometry. In the Structures III and IV, the C of the $UC_2$ bonds with the C of graphene, making the graphene partially defected in the adsorption [27,53].

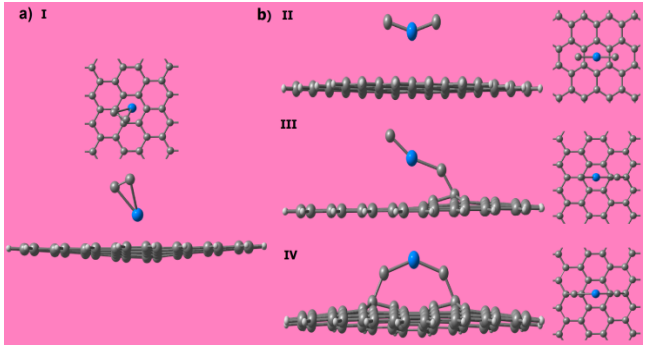

**Figure 1.** Optimized geometries for the graphene-supported $UC_2$ complexes (top and side views). a) $UC_2$ ($C_{2v}$-symmetric)/graphene, denoted as Structure I. b) $UC_2$ (linear)/graphene, denoted as Structures II, III and IV. Throughout the remaining part of the article, the C on the left and right sides of the $UC_2$ of each system in Figure 1 are called as $C_{1st}$ and $C_{2nd}$, respectively.

The obtained relative energies $\Delta E$, geometrical parameters, and charge transfer of these four systems are included in Table 1. To compare the interaction of these systems, in Table 1 we also present the calculated adsorption $E_{ads}$, interaction $E_{int}$ and deformation $E_{def}$ energies defined as

$$E_{ads/int} = (E_{iso/relaxed}^{UC_2} + E_{iso/relaxed}^{graphene}) - E_{UC_2+graphene} \quad (1)$$

$$E_{def}^{UC_2+graphene} = E_{def}^{UC_2} + E_{def}^{graphene} = E_{int} - E_{ads}$$
$$= (E_{relaxed}^{UC_2} - E_{iso}^{UC_2}) + (E_{relaxed}^{graphene} - E_{iso}^{graphene}) \quad (2)$$

where $E_{UC_2+graphene}$ is the total energy of the $UC_2$ molecule adsorbed on the graphene surface, $E_{iso}^{UC_2}$ and $E_{iso}^{graphene}$ represent the total energies of the isolated $UC_2$ and graphene sheet, respectively, and $E_{relaxed}^{UC_2} / E_{relaxed}^{graphene}$ denotes the total energy of $UC_2$ or graphene in its relaxed geometry, with both the $UC_2$ and graphene being in the same atomic configurations as in the relaxed $UC_2$+graphene system.

The interaction energy $E_{int}$ represents a measure of the strength of the chemical interaction between the $UC_2$ molecule and the graphene surface; a positive $E_{int}$ means an attractive interaction, while the adsorption energy $E_{ads}$ encompasses not only the energy gained due to the chemical interaction $E_{int}$ but also the energy paid to deform the molecules and surface from their ideal configurations to the relaxed molecule-surface system, which are occurred during the adsorption process.

As can be seen from Figure 1 and Table 1, for all the four adsorption structures, the deformations of graphene surface are more than 0.1 Å; the charge transfer implies electrostatic attraction; and the interaction energies and the adsorption energies are relatively higher than the general intermolecular physisorption energy. Similar to previous reports about metal chemisorption on graphene, [54] these structures might present some chemisorption characteristics which will be examined further in the following. The $UC_2$ ($C_{2v}$-symmetric)/graphene (i.e. Structure I) is the energetically most favorable adsorption structure with a high adsorption energy (up to 2.27 eV), where the U donates considerable amount of electrons (1.133 e). Despite the relatively high energy, the systems of the linear CUC adsorbed on graphene can be formed under the extreme environment in the core of nuclear reactor and only the linear CUC was so far observed in experiment [34]. Therefore, the interaction between this linear CUC and graphene requires further studies. Among the three structures, the Structure IV is the most representative, involving a strong chemical interaction with the highest interaction energy (6.05 eV). The C atoms at the two ends of the $UC_2$ both bond with the graphene, inducing double $sp^3$-hybridization in graphene and then protruding the adsorption region of graphene to become a defect with a maximum deformation of about -0.54 Å. In addition to the main results of the Structures I and IV we present here, we also provide the results of Structures II and III in part 4 of the Supporting Information.

To facilitate analysis of the electron density properties, the side view color-filled maps of electron density (Figure 2) and the contour maps of charge density deformation (Figure 3) between the adsorption structure and the non-interaction fragments (in the same position) were calculated from the molecular orbitals using a Multiwfn program [55].

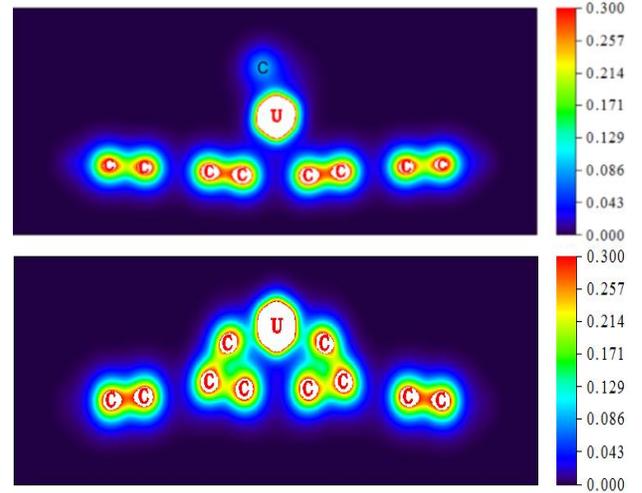

**Figure 2.** Color-filled maps of electron density of Structures I and IV (sectional views). In Structure I, the black "C" indicates that this carbon atom is not on the two-dimensional cutting plane.

Further analysis of the charge density deformation reveals strong chemical interaction as is visualized in Figure 3 which shows intense charge transfer from the $UC_2$ to the graphene near the adsorption site. It is generally believed that for covalent bonding due to the overlap of molecular orbitals electrons would accumulate at the center of the bond. The electron density increases at the intermediate region between the U and the beneath hexagonal ring in both Structures I and IV, as indicated with the black solid lines. Moreover, the charge accumulation between the C at either side of the $UC_2$ and the C of graphene is obvious in Structure IV. All the above reflect the strong interaction between the $UC_2$ molecule and graphene.

**Table 1.** Relative ($\Delta E$), adsorption ($E_{ads}$), interaction ($E_{int}$) and deformation ($E_{def}$) energies (eV) for the ground state of each adsorption structure. Adsorption distance ($D$, Å)[a], graphene surface deformation ($\Delta Z$, Å)[b] and charge transfer ($\Delta Q$, e) are also listed.

|     | $\Delta E$ | $E_{ads}$ | $E_{int}$ | $E_{def}^{UC_2}$ | $E_{def}^{graphene}$ | $D$ | $\Delta Z$ | $\Delta Q$ (U) | $\Delta Q$ ($C_{1st}$) | $\Delta Q$ ($C_{2nd}$) |
| --- | --- | --- | --- | --- | --- | --- | --- | --- | --- | --- |
| I   | 0    | 2.27 | 2.35 | -0.08 | 0.16 | 2.40 | 0.52  | 1.133 | -0.046 | -0.023 |
| II  | 3.69 | 1.25 | 1.46 | 0.16  | 0.05 | 2.57 | 0.24  | 0.418 | 0.007  | 0.007  |
| III | 3.48 | 1.47 | 4.21 | 1.53  | 1.21 | 2.41 | -0.14 | 0.977 | -0.072 | 0.109  |
| IV  | 4.02 | 0.93 | 6.05 | 2.75  | 2.38 | 2.29 | -0.54 | 1.116 | 0.213  | 0.213  |

[a] $D = \overline{Z}(U) - \overline{Z}(6C)$, the adsorption distance ($D$) of the $UC_2$ from the graphene surface, where $\overline{Z}(6C)$ is the average $Z$ coordinates of the C hexagonal ring beneath the U atom. [b] $\Delta Z = \overline{Z}(24H) - \overline{Z}(6C)$, the graphene surface deformation ($\Delta Z$), defined as the difference in average of the $Z$ value of 24 H atoms around the graphene and the nearest six center C atoms beneath the U atom.

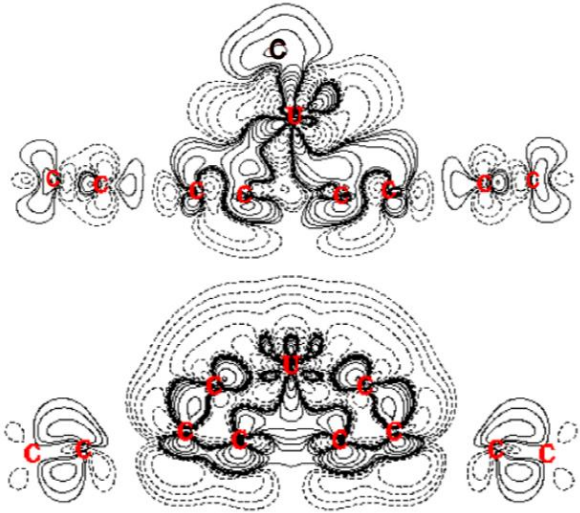

**Figure 3.** Contour maps of charge density deformation of Structures I and IV (sectional views). The black solid lines and dashed lines show the charge accumulation and depletion, respectively. In Structure I, the black "C" indicates that this carbon atom is not on the two-dimensional cutting plane. (The corresponding isosurfaces are given in part 3 of the Supporting Information)

The aforementioned analyses uncover a strong interaction rather than a weak adsorption for the $UC_2$ on the graphene surface, with intense charge transfer. It indicates the necessity for us to understand the origin of the interaction by orbital analysis. Thus, the frontier molecular orbitals (only for the adsorption central region) active valence electrons of Structures I and IV are displayed in Figures S6 and S8. The Natural Atomic Orbitals (NAO) approach in Natural Bond Orbitals (NBO) analysis [56] was used to quantitatively analyze the contribution from atomic orbitals to these frontier molecular orbitals based on the eigenstates. Specifically, we focus on the 5f and 6d orbitals of the U and the 2p orbitals of C at both sides of the $UC_2$. Their contribution percentages are listed in Tables 2 and 3.

These molecular orbital diagrams show that there are orbital overlaps between the U and the beneath C of graphene in Structures I and IV, well proving the existence of chemisorption. Moreover, as shown in HOMOα, HOMOα-2 and HOMOα-4 of Figure S8, there are not only orbital overlaps between the U and the C of graphene but also overlaps between the C at either side of the $UC_2$ and the C of graphene, forming double $sp^3$-hybridization in Structure IV.

Results in Tables 1-3 reveal that the interaction between the $UC_2$ and graphene is mainly from the contribution of 5f (the main source) and 6d orbitals of the U and the hexagonal ring of graphene for Structure I. In Structure IV, the $UC_2$ also bonds with graphene via 2p orbitals of the two C atoms. The orbital overlaps with much lower energy than the HOMO region show strong orbital interactions between the graphene and the $UC_2$, and the frontier orbitals of the system are not localized on the chemical bond. Moreover, the charge transfer implies electrostatic interactions, especially in Structure I; the frontier orbital with U 7s character disappears, indicating that the U atom donates electrons mainly to the beneath C atoms of graphene, leading to a donor-acceptor interaction.[57] As a result of such covalent and electrostatic interactions, the adsorption distances ($D$) of the U are all shorter than the U-C distances (avg. 2.93 Å) observed in the $U^{3+}$ compound, $C_6Me_6U(BH_4)_3$.[58]

The adsorption of the $UC_2$ changes the electronic energy levels of the systems, as shown clearly in the density-of-states (DOS) of the Structures I and IV given in Figure 4. The effects of adsorption can be also seen in the local DOS (LDOS) of the two fragments, the $UC_2$ and the graphene.

By comparing the DOS of the isolated intrinsic graphene and the graphene in the adsorption system, we observed that the interaction between the $UC_2$ and graphene changed the electronic structure of graphene significantly. Specifically, the DOS of α and β electrons are no longer symmetrical near the Fermi level, inducing obvious spin polarization carried by the $UC_2$. Similar effect has been reported in previous study on adsorption defects of graphene [59]. Comparing the TDOS of these two kinds of structures reveals more remarkable spin polarization in Structure I. The adsorption of the $UC_2$ induced significant change in the electronic structures of the graphene provides further evidence of the strong chemisorption.

The peaks in LDOS of Structures I and IV indicate that the orbitals of the $UC_2$ and graphene both contribute to the molecular orbitals of the systems at certain energy levels, which further reflects the adsorption interaction. As mentioned above, there are overlaps of orbitals, HOMOα-4, HOMOα-5 for Structure I, between the $UC_2$ and graphene, shown in Figure S6, and HOMOα-2, HOMOα-4 for Structure IV in Figure S8. They are also shown as the peaks in LDOS near the energy levels at -5.83 eV for Structure I, and at -4.94 eV and -5.31 eV for Structure IV.

**Table 2.** Results of population analysis of frontier molecular orbitals at the adsorption region of Structure I. [a]

| Structure I | $U_{121}$ | | | $C_{1st}$ | | $C_{2nd}$ | | ring[b] |
|---|---|---|---|---|---|---|---|---|
| | total | 5f | 6d | total | 2p | total | 2p | total |
| **HOMOα** | 4.27% | 1.43% | 1.55% | 0.05% | 0.04% | 0.01% | 0.01% | 2.30% |
| HOMOα-1 | 10.41% | 10.24% | 0.13% | 0.27% | 0.10% | 0.57% | 0.45% | 2.24% |
| HOMOα-2 | 4.70% | 4.45% | 0.22% | 0.12% | 0.11% | 0.07% | 0.07% | 2.79% |
| HOMOα-3 | 83.60% | 77.85% | 5.48% | 4.36% | 3.06% | 2.47% | 1.99% | 1.15% |
| **HOMOα-4** | 81.58% | 78.50% | 2.12% | 1.48% | 1.48% | 3.12% | 2.52% | 1.44% |
| **HOMOα-5** | 84.11% | 83.23% | 0.53% | 1.87% | 1.50% | 0.64% | 0.64% | 1.00% |
| HOMOα-6 | 1.57% | 1.56% | 0.01% | 0.03% | 0.02% | 0.01% | 0.01% | 0.04% |
| **HOMOα-7** | 1.28% | 0.95% | 0.30% | 0.03% | 0.03% | 0.09% | 0.08% | 4.39% |
| **HOMOα-8** | 1.82% | 1.12% | 0.29% | 0.26% | 0.24% | 0.27% | 0.24% | 4.86% |

[a] The bold fonts represent the overlap orbitals between the U and graphene. [b] The "ring" represents the hexagonal ring C beneath the U.

**Table 3.** Results of population analysis of frontier molecular orbitals at the adsorption region of Structure IV. [a]

| Structure IV | $U_{1st}$ | | | $C_{1st}$ | | $C_{2nd}$ | | ring[b] |
|---|---|---|---|---|---|---|---|---|
| | total | 5f | 6d | total | 2p | total | 2p | total |
| **HOMOα** | 67.89% | 60.25% | 6.79% | 1.48% | 1.48% | 1.49% | 1.48% | 13.20% |
| HOMOα-1 | 53.01% | 50.93% | 0.66% | 19.30% | 19.30% | 19.18% | 19.18% | 1.34% |
| **HOMOα-2** | 5.72% | 4.06% | 1.17% | 4.01% | 4.01% | 4.00% | 3.98% | 4.08% |
| HOMOα-3 | 4.63% | 0.18% | 4.44% | 5.92% | 5.92% | 5.99% | 5.99% | 0.94% |
| <u>HOMOα-4</u> | 2.69% | 2.47% | 0.11% | 0.79% | 0.79% | 0.79% | 0.79% | 1.73% |
| HOMOα-5 | 90.56% | 88.96% | 0.84% | 0.19% | 0.19% | 0.19% | 0.19% | 2.30% |
| HOMOα-6 | 22.60% | 2.21% | 20.38% | 19.84% | 19.84% | 19.90% | 19.90% | 6.03% |
| HOMOα-7 | 0.70% | 0.67% | 0.00% | 0.48% | 0.48% | 0.48% | 0.48% | 0.08% |
| **HOMOα-8** | 4.69% | 4.20% | 0.41% | 0.18% | 0.18% | 0.18% | 0.18% | 4.15% |

[a] The bold fonts represent the overlap orbitals between the U and graphene; the underlined fonts represent the overlap orbitals between the C of $UC_2$ and the C of graphene. [b] The "ring" represents the the hexagonal ring C beneath the U.

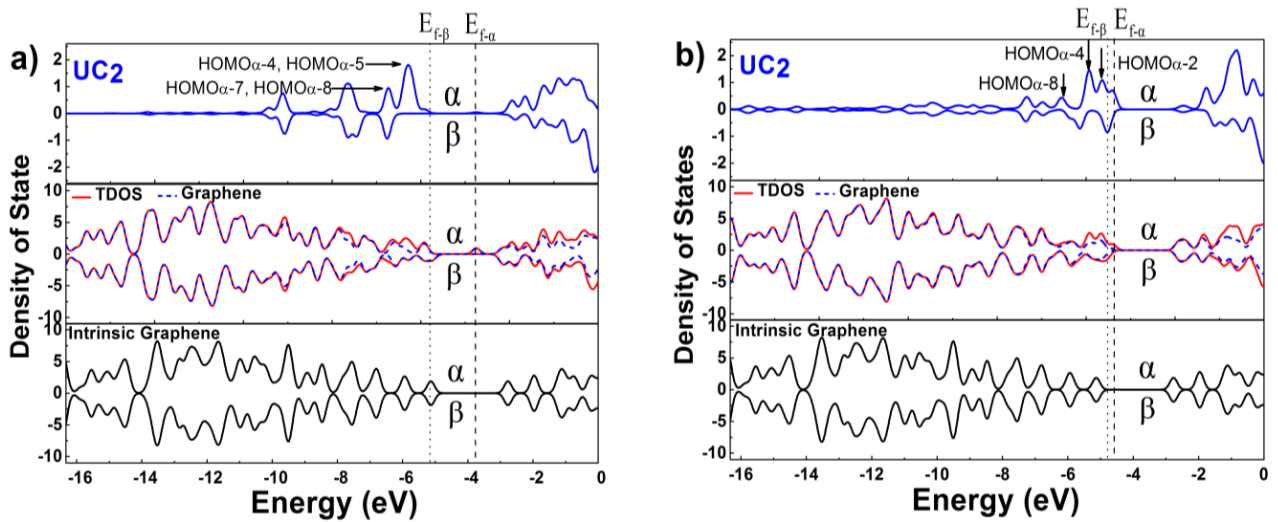

**Figure 4.** DOS of Structure I (a) and Structure IV (b). The $E_{f-α}$ and $E_{f-β}$ indicate the Fermi level of α and β electrons of each adsorption system, respectively.

It is interesting to understand the adsorption behaviors in different systems. For Structures I and II, obviously they both involve interactions between the graphene and the U atom, and the deformation energies are negligible (Table 1). Consequently, the difference in the adsorption energies origins from the chemical interaction between the U atom and graphene (i.e. $E_{int}$). In Structure I, the charge transfer is more intense and the adsorption distance is shorter than that in Structure II, mainly because that there are more active U 5f orbitals in symmetric triangular $UC_2$ than in CUC (as already discussed in the last paragraph of the INTRODUCTION). Moreover, the unsaturated and diffuse 5f orbitals of U could enhance the π bonding. [60] Therefore, we suggest that for the symmetric triangular $UC_2$ and compared with the CUC, the 5f electrons of U can participate in bonding to a greater degree and generate stronger interaction with the graphene which forms delocalized π bond by $sp^2$-hybridization, giving rise to a higher interaction energy. In Figure S7, there is only HOMOα-4 with obvious overlap between the CUC and graphene in Structure II. Similar results are also shown in overlap population DOS (OPDOS). The bonding contribution area at the occupied states in Structure I is larger than that in Structure II (see part 6 in the Supporting Information). For the systems of CUC adsorbed on graphene, the adsorption distances and charge transfer are roughly linear correlated to the $E_{int}$ in Structures II, III and IV. For Structure IV, the CUC strongly bonds with the graphene, resulting in the highest interaction energy (6.05 eV) with a double $sp^3$-hybridization. The interaction energy, however, mostly compensates with the large energy required to distort the $UC_2$ molecule and the graphene surface upon adsorption (Table 1). The relative high energy and low adsorption energy of Structure IV indicate an energetically unfavorable system.

In consequence, we find clear evidence for strong adsorption characteristic on the electronic structure. This characteristic that differs from weak intermolecular adsorption can be clarified by a feature of collective vibration and stretching vibration between the $UC_2$ and graphene on the vibration spectrum. Therefore, we make a further analysis of the vibration spectrum characteristics of the systems.

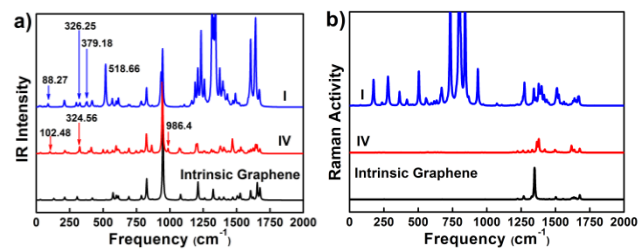

**Figure 5.** The infrared (a) and Raman (b) vibrational spectra of Structures I and IV. The arrows indicate the peaks generated by the interaction between the $UC_2$ and graphene.

Figure 5 presents the infrared and Raman vibrational spectra obtained from our theoretical calculations for Structures I and IV. In general, the higher in adsorption energy the system is, the greater change exists in its infrared and Raman activities and the more complex the vibration mode is. Because of the existence of strong interaction between the $UC_2$ and graphene, and vibration modes at the frequency above 1500 $cm^{-1}$ belong to the vibration between the small molecules of C and H, we examined the collective vibration and stretching vibration modes below 1500 $cm^{-1}$. We found that there existed such infrared active modes between the U and the hexagonal ring of graphene at 88.27, 326.25 and 379.18 $cm^{-1}$ in Structure I and 102.48, 324.56 $cm^{-1}$ in Structure IV (see Figure S10). The result further reveals the strong interaction in the systems. Specially, there are stretching vibration modes between the C at both sides of the $UC_2$ and the C of graphene surface at 986.40 $cm^{-1}$ in Structure IV (see Figure S11), reflecting the strong bonding which transforms the hybridization mode from $sp^2$ to $sp^3$ upon adsorption. Interestingly, the characteristic peaks of these vibration modes described above are infrared active only and can be found in the infrared spectrum of vibration but not in the Raman spectrum. There is also an intense peak at 518.66 $cm^{-1}$ in Structure I which indicates the symmetric stretching mode of triangular $UC_2$, in agreement with Wang and Andrews et al.'s calculations which showed that vibrations of the triangular $UC_2$ should fall around 500 $cm^{-1}$ (see this mode in Figure S1 (b) of the Supporting Information). However, the triangular $UC_2$ isomer was not detected here in their experiment [34]. This vibrational mode also appears in our adsorption system, which may be a fingerprint of the U-$C_2$ bonding to guide future experimental characterization.

## ■ CONCLUSIONS

In summary, using hybrid DFT-PBE level first-principles calculations, we investigated the adsorption of two $UC_2$ isomers (linear CUC and symmetric triangular structures) on graphene surface. We found strong chemisorption in the four stable adsorption structures, as evidenced in the charge transfers and orbital overlaps, as well as in the existence of collective vibration and stretching vibration between the $UC_2$ and graphene. Although the Structures III and IV are energetically less favorable, it is possible that the CUC can adsorb onto graphene surface and form the structure with $sp^3$ hybridization under the extreme environment in the core of nuclear reactor. The Structure I formed by adsorption of the symmetric triangular $UC_2$ onto graphene is energetically the most favorable, with mainly U atom participating in the adsorption. Bonding analysis highlights the dominant role of 5f valence electrons of U in the interaction.

Due to the adsorption of the $UC_2$ onto graphene, the structure and electronic properties of graphene are changed. It has been reported that the adsorption of foreign matter can affect the thermal conductivity and reduce the mechanical strength of graphene. [61] The microstructure change can affect the anisotropy, thermal conductivity and mechanical property of PyC of coated particle nuclear fuel. [62] Thus, our study would be highly relevant to the burn-ups improvement and coating failure reduction. These property changes induced by the adsorption of the $UC_2$ should be taken into account in the fuel design and subsequent reactor safety assessments.

The research of computational chemistry of actinides is a frontier subject nowadays and has received much attention. [12] The unique bonding properties of 5f electrons make actinide elements and their compounds presenting interesting nature. [63] Our calculations on the interaction of the $UC_2$ and graphene show the prominent chemisorption behavior caused by the delocalization of 5f electrons, which is important for understanding the role of 5f electrons of actinide elements in bonding. It can be predicted that graphite will adsorb and retain gaseous $UC_2$ molecules as possible fission products without being significantly damaged in Gen-IV nuclear reactor. In general nuclear system, the structural integrity and the deformation under stress of nuclear graphite components are essential elements of fuel design and safety assessments. [64,65] Our research predicts the changes and their mechanism of nuclear graphite, and provides theoretical support to the current and future advanced nuclear technology. The nuclear fuel reprocessing and nuclear waste disposal by using adsorption of carbon materials to actinides and related compounds have received extensive attention.[66] At present, there has been an experiment reporting the absorption between depleted uranium and graphene.[10] Our work also provides a reference for related research and applications.

## AUTHOR INFORMATION
### Corresponding Author
wangzg@jlu.edu.cn; huaiping@sinap.ac.cn; aprqz@cityu.edu.hk

## ACKNOWLEDGMENT
The work described in this paper was supported by the Strategically Leading Program of the Chinese Academy of Sciences (Grant no. XDA02040100), and City University of Hong Kong (Project No. 7002749). Z.W. acknowledges the High Performance Computing Center (HPCC) of Jilin University and NSFC Nos. 11374004 and 11004076.

## REFERENCES
(1) Geim, A. K. *Science* **2009**, *324*, 1530
(2) Simnad, M. T. *Nuclear Reactor Materials and Fuels*; Encyclopedia of Physical Science and Technology, Academic Press, New York, 1987.
(3) Cahn, R. W. *J. Nucl. Mater.* **1983**, *114*, 116.
(4) B, E. M.; A. *J. Phys: Conf. Ser.* **2012**, *371*, 012017.
(5) Martin, D. G. *Nucl. Eng. Des.* **2002**, *213*, 241.
(6) *High Temperature Gas Cooled Reactor Fuels and Materials*; IAEA TECDOC (CD-ROM) 1645 ed.; International Atomic Energy Agency (IAEA): Vienna, 2010.
(7) Campbell, F. R. *Interaction Between Graphite and UO2 in Operating Nuclear Fuel Elements*; Atomic Energy of Canada Limited, Fuel Materials Branch, Chalk River Nuclear Laboratories: Chalk River, Ontario, Canada, 1976.
(8) Yang, X.; Feng, S.; Zhou, X.; Xu, H.; Sham, T. K. *J. Phys. Chem. A* **2012**, *116*, 985.
(9) Sheil, R. J.; Evans, R. B.; Watson, G. M. *Oak Ridge National Lab., Tenn.* **1959**.
(10) Cerefice, G. S.; Schmidt, G. H.; Keith, C. In *2012 GSA Annual Meeting in Charlotte* **2012**.
(11) Maeda, K.; Sasaki, S.; Kato, M.; Kihara, Y. *J. Nucl. Mater.* **2009**, *389*, 78.
(12) Wang, D.; van Gunsteren, W. F.; Chai, Z. *Chem. Soc. Rev.* **2012**, *41*, 5836.
(13) Srivastava, D.; Garg, S. P.; Goswami, G. L. *J. Nucl. Mater.* **1989**, *161*, 44.
(14) Petti, D.; Crawford, D.; Chauvin, N. *MRS Bull.* **2009**, *34*, 40.
(15) Utton, C. A.; De Bruycker, F.; Boboridis, K.; Jardin, R.; Noel, H.; Guéneau, C.; Manara, D. *J. Nucl. Mater.* **2009**, *385*, 443.
(16) Hunt, R. D.; Lindemer, T. B.; Hu, M. Z.; Del Cul, G. D.; Collins, J. L. *Radiochim. Acta* **2007**, *95*, 225.
(17) Bowman, A. L.; Arnold, G. P.; Witteman, W. G.; Wallace, T. C.; Nereson, N. G. *Acta Crystallogr.* **1966**, *21*, 670.
(18) Atoji, M. *J. Chem. Phys.* **1967**, *47*, 1188.
(19) Zalazar, M. F.; Rayón, V. M.; Largo, A. *J. Phys. Chem. A* **2012**, *116*, 2972.
(20) Pogány, P.; Kovács, A.; Varga, Z.; Bickelhaupt, F. M.; Konings, R. J. M. *J. Phys. Chem. A* **2011**, *116*, 747.
(21) Datta, B. P.; Sant, V. L.; Raman, V. A.; Subbanna, C. S.; Jain, H. C. *Int. J. Mass Spectrom. Ion Processes* **1989**, *91*, 241.
(22) Norman, J. H.; Winchell, P. *J. Phys. Chem.* **1964**, *68*, 3802.
(23) Miller, G. K.; Petti, D. A.; Varacalle, D. J.; Maki, J. T. *J. Nucl. Mater.* **2001**, *295*, 205.
(24) Kang, S. G.; Zhou, G.; Yang, P.; Liu, Y.; Sun, B.; Huynh, T.; Meng, H.; Zhao, L.; Xing, G.; Chen, C.; Zhao, Y.; Zhou, R. *PNAS* **2012**, *109*, 15431.
(25) Diener, M. D.; Smith, C. A.; Veirs, D. K. *Chem. Mater.* **1997**, *9*, 1773.
(26) Jackson, K.; Kaxiras, E.; Pederson, M. R. *J. Phys. Chem.* **1994**, *98*, 7805.
(27) Dai, X.; Cheng, C.; Zhang, W.; Xin, M.; Huai, P.; Zhang, R.; Wang, Z. *Sci. Rep.* **2013**, *3*, 1341.
(28) Wu, X.; Lu, X. *J. Am. Chem. Soc.* **2007**, *129*, 2171.
(29) Dunk, P. W.; Kaiser, N. K.; Mulet-Gas, M.; Rodríguez-Fortea, A.; Poblet, J. M.; Shinohara, H.; Hendrickson, C. L.; Marshall, A. G.; Kroto, H. W. *J. Am. Chem. Soc.* **2012**, *134*, 9380.
(30) Liu, X.; Li, L.; Liu, B.; Wang, D.; Zhao, Y.; Gao, X. *J. Phys. Chem. A* **2012**, *116*, 11651.
(31) Guo, T.; Diener, M. D.; Chai, Y.; Alford, M. J.; Haufler, R. E.; McClure, S. M.; Ohno, T.; Weaver, J. H.; Scuseria, G. E.; Smalley, R. E. *Science* **1992**, *257*, 1661
(32) Londono-Hurtado, A.; Szlufarska, I.; Morgan, D. *J. Nucl. Mater.* **2013**, *437*, 389.
(33) Wang, X.; Andrews, L.; Malmqvist, P. Å.; Roos, B. O.; Gonçalves, A. P.; Pereira, C. C. L.; Marçalo, J.; Godart, C.; Villeroy, B. *J. Am. Chem. Soc.* **2010**, *132*, 8484.
(34) Wang, X.; Andrews, L.; Ma, D.; Gagliardi, L.; Goncalves, A. P.; Pereira, C. C. L.; Marcalo, J.; Godart, C.; Villeroy, B. *J. Chem. Phys.* **2011**, *134*, 244313.
(35) Ross, B. O. In *Adv. Chem. Phys.*; Lawley, K. P., Ed.; John Wiley & Sons: Chichester, UK, 1987; Vol. 69, p 399.
(36) Parr, R. G. *Annu. Rev. Phys. Chem.* **1983**, *34*, 631.
(37) Dolg, M. In *Theoretical and Computational Chemistry*; Peter, S., Ed.; Elsevier: 2002; Vol. 11, p 793.
(38) Christiansen, P. A.; Ermler, W. C.; Pitzer, K. S. *Annu. Rev. Phys. Chem.* **1985**, *36*, 407.
(39) Pyykko, P. *Chem. Rev.* **1988**, *88*, 563.
(40) Ernzerhof, M.; Scuseria, G. E. *J. Chem. Phys.* **1999**, *110*, 5029.
(41) Perdew, J. P.; Burke, K.; Ernzerhof, M. *Phys. Rev. Lett.* **1996**, *77*, 3865.
(42) Adamo, C.; Barone, V. *J. Chem. Phys.* **1999**, *110*, 6158.
(43) Andersson, K.; Malmqvist, P. A.; Roos, B. O.; Sadlej, A. J.; Wolinski, K. *J. Phys. Chem.* **1990**, *94*, 5483.
(44) Andersson, K.; Malmqvist, P. A.; Roos, B. O. *J. Chem. Phys.* **1992**, *96*, 1218.
(45) Kuchle, W.; Dolg, M.; Stoll, H.; Preuss, H. *J. Chem. Phys.* **1994**, *100*, 7535.
(46) Cao, X.; Dolg, M.; Stoll, H. *J. Chem. Phys.* **2003**, *118*, 487.
(47) Binkley, J. S.; Pople, J. A.; Hehre, W. J. *J. Am. Chem. Soc.* **1980**, *102*, 939.
(48) Zhang, R.; Lifshitz, C. *J. Phys. Chem.* **1996**, *100*, 960.
(49) Cong, Y.; Yang, Z.-Z.; Wang, C.-S.; Liu, X.-C.; Bao, X.-H. *Chem. Phys. Lett.* **2002**, *357*, 59.
(50) Forte, G.; Grassi, A.; Lombardo, G. M.; La Magna, A.; Angilella, G. G. N.; Pucci, R.; Vilardi, R. *Phys. Lett. A* **2008**, *372*, 6168.
(51) Rudenko, A. N.; Keil, F. J.; Katsnelson, M. I.; Lichtenstein, A. I. *Phys. Rev. B* **2012**, *86*, 075422.
(52) Frisch, M. J.; et al.; Gaussian 09, revision D. 01, Gaussian, Inc., Wallingford CT, **2013**. See the Supporting Information for full reference.
(53) Xin, M.; Wang, F.; Meng, Y.; Tian, C.; Jin, M.; Wang, Z.; Zhang, R. *J. Phys. Chem. C* **2011**, *116*, 292.
(54) Hu, L.; Hu, X.; Wu, X.; Du, C.; Dai, Y.; Deng, J. *Physica B* **2010**, *405*, 3337.
(55) Lu, T.; Chen, F. *J. Comput. Chem.* **2012**, *33*, 580.
(56) Reed, A. E.; Weinhold, F. *J. Chem. Phys.* **1983**, *78*, 4066.
(57) Bent, H. A. *Chem. Rev.* **1968**, *68*, 587.
(58) Baudry, D.; Bulot, E.; Charpin, P.; Ephritikhine, M.; Lance, M.; Nierlich, M.; Vigner, J. *J. Organomet. Chem.* **1989**, *371*, 155.
(59) Liu, H. Y.; Hou, Z. F.; Hu, C. H.; Yang, Y.; Zhu, Z. Z. *J. Phys. Chem. C* **2012**, *116*, 18193.
(60) Morss, L. R.; Edelstein, N. M.; Fuger, J.; Katz, J. J. *The chemistry of the actinide and transactinide elements*; Springer, 2006.
(61) Banhart, F.; Kotakoski, J.; Krasheninnikov, A. V. *ACS Nano* **2010**, *5*, 26.
(62) López-Honorato, E.; Meadows, P. J.; Xiao, P.; Marsh, G.; Abram, T. J. *Nucl. Eng. Des.* **2008**, *238*, 3121.
(63) Moore, K. T.; van der Laan, G. *Rev. Mod. Phys.* **2009**, *81*, 235.
(64) Tsang, D. K. L.; Marsden, B. J. *J. Nucl. Mater.* **2006**, *350*, 208.
(65) Tsang, D. K. L.; Marsden, B. J. *Nucl. Eng. Des.* **2007**, *237*, 897.
(66) Deb, A.; Ilaiyaraja, P.; Ponraju, D.; Venkatraman, B. *J. Radioanal. Nucl. Chem.* **2012**, *291*, 877.

# Supporting Information

# Strong Adsorption between Uranium Dicarbide and Graphene Surface Induced by f Electrons

Jie Han,[†] Xing Dai,[†] Cheng Cheng,[‡] Minsi Xin,[†,§] Zhigang Wang,*[,†,§] Ping Huai,*[,‡] and Ruiqin Zhang*[,§,#]

[†]Institute of Atomic and Molecular Physics, Jilin University, Changchun 130012, P. R. China

[‡]Shanghai Institute of Applied Physics, Chinese Academy of Sciences, Shanghai 201800, P. R. China

[§]Beijing Computational Science Research Center, Beijing, P.R. China

[#]Department of Physics and Materials Science and Centre for Functional Photonics (CFP), City University of Hong Kong, Hong Kong SAR, P. R. China

**Contents**

# Part 1. Electronic structures of UC$_2$ and validation of calculation method

Table S1. Calculated relative energies $\Delta E$ (eV) for linear CUC (D$_{\infty h}$) at various levels of theory.

| linear CUC | Spin multiplicity | $\Delta E$ |
|---|---|---|
| PBE0 | Singlet | 0.09 |
| | **Triplet** | **0.00** |
| | Quintet | 0.30 |
| B3LYP | Singlet | 1.27 |
| | Triplet | 0.12 |
| | **Quintet** | **0.00** |
| PBEPBE | Singlet | 0.66 |
| | Triplet | 0.58 |
| | **Quintet** | **0.00** |
| PW91PW91 | Singlet | 1.01 |
| | Triplet | 0.33 |
| | **Quintet** | **0.00** |
| BP86 | Singlet | 1.34 |
| | Triplet | 0.14 |
| | **Quintet** | **0.00** |

*Note: Basis set for U atom：Stuttgart Relativistic Large Core ECP (78 core electronics) [S1], and for C and H atoms: 3-21G; those in bold represent the ground state at various functionals. PBE0 and PBEPBE denote the hybrid and pure PBE functional of DFT, respectively.

As seen in Table S1, at the Stuttgart RECP (78 core electronics) level, only by using PBE0 functional can we get the result that the ground state of linear CUC is Triplet which is in agreement with former researches based on calculations using SO-CASPT2 [20] and DFT [34]. In the next step, we continue to calculate the symmetric triangular UC$_2$ with PBE0 functional to test the reliability of the PBE0 functional.

Table S2. Calculated relative energies $\Delta E$ (eV) and geometrical parameters (Å, deg) for linear CUC (D$_{\infty h}$), sym $\Delta$U-C$_2$ (C$_{2v}$) by using the PBE0 functional.

| Structure | Spin multiplicity | $\Delta E$ | d(UC) | d(CC) | ∠CUC |
|---|---|---|---|---|---|
| linear CUC | Singlet | 0.09 | 1.811 | | 180.0 |
| | **Triplet** | **0.00** | **1.852** | | **180.0** |
| | Quintet | 0.30 | 1.932 | | 180.0 |
| sym$\Delta$ U-C$_2$ | Triplet | 1.39 | 2.302 | 1.280 | 32.3 |
| | **Quintet** | **0.00** | **2.326** | **1.285** | **32.1** |
| | Septet | 2.38 | 2.624 | 1.279 | 28.2 |

*Note: Basis set for U atom: Stuttgart Relativistic Large Core ECP (78 core electronics), and for C and H atoms: 3-21G. Those in bold are the ground state.

As seen in Table S2, at the Stuttgart RECP (78 core electronics) level, by using PBE0 functional we can get the results that the ground state of the linear CUC is Triplet and the symmetric triangular $UC_2$ is Quintet, in agreement with former researches based on calculations using SO-CASPT2 [20] and DFT [34]. However, the results of geometry has a little deviation, so we used ECP60MWB_SEG (60 core electronics) for further calculations.

Table S3. Calculated relative energies $\Delta E$[a] (eV) and geometrical parameters (Å,deg), harmonic frequencies, and infrared intensities of the ground states for CUC ($D_{\infty h}$) and sym$\Delta$ U-$C_2$ ($C_{2v}$) at various levels of theory.

| Structure | Method | $\Delta E$ | $d$(UC) | $d$(CC) | ∠CUC | M[f] | Vibrational Frequencies |
|---|---|---|---|---|---|---|---|
| CUC ($D_{\infty h}$) linear Triplet | SO-CASPT2[b] | 2.95 | 1.825 | | 180.0 | 3 | |
| | BPW91[c] | 2.95 | 1.846 | | 180.0 | 3 | 54(89×2), 891(0), 952(228) |
| | B3LYP[d] | 3.64 | 1.834 | | 180.0 | 3 | 110(88), 918(0), 976(252) |
| | PBE0/3-21G[e] | 2.69 | 1.837 | | 180.0 | 3 | 157(75×2), 936(0), 999(182) |
| | PBE0/6-31G[e] | 2.98 | 1.838 | | 180.0 | 3 | 131(78), 941(0), 1000(197) |
| | M06L/6-31G[e] | 2.53 | 1.849 | | 180.0 | 3 | |
| U-$C_2$ ($C_{2v}$) sym Δ Quintet | SO-CASPT2[b] | 0 | 2.257 | 1.271 | 32.7 | 5 | |
| | BPW91[c] | 0 | 2.243 | 1.281 | | 5 | 336(2), 466(50), 1698(25) |
| | B3LYP[d] | 0 | 2.291 | 1.261 | 32.0 | 5 | 238(1), 508(130), 1826(2) |
| | PBE0/3-21G[e] | 0 | 2.279 | 1.283 | 32.7 | 5 | 235(0), 517(119), 1770(1) |
| | PBE0/6-31G[e] | 0 | 2.269 | 1.285 | 32.9 | 5 | 266(1), 512(111), 1792(1) |
| | M06L/6-31G[e] | 0 | 2.289 | 1.290 | 32.7 | 5 | 234(0), 497(114), 1736(2) |

[a] $\Delta E$ refers to the difference with respect to the ground-state. [b] Ref. [20]. [c] Ref. [34]. [d] Ref. [19]. [e] tested methods used in our computation. [f] Spin multiplicity.

Table S4. Calculated Mulliken charge population at PBE0/3-21G and PBE0/6-31G levels for $UC_2$.

| Structures | Method | Mulliken charge population | |
|---|---|---|---|
| | | U | C |
| CUC ($D_{\infty h}$, Triplet) | **PBE0/3-21G** | 0.409 | -0.235 |
| | PBE0/6-31G | -0.208 | 0.104 |
| U-$C_2$ ($C_{2v}$, Quintet) | **PBE0/3-21G** | 0.542 | -0.271 |
| | PBE0/6-31G | 0.240 | -0.120 |

As seen in Table S3, by using DFT-PBE0 with ECP60MWB_SEG and 3-21G, we obtained results of the $UC_2$ which are basically the same as previous theoretical researches. And in Table S4, the calculated Mulliken charge population of symmetric triangular $UC_2$ at PBE0/3-21G level (+0.542 for U) is in line with the charge distribution assessed in reference [20] (+0.5 for U). This is also consistent with the molecular orbital analysis of symmetric triangular $UC_2$ reported in reference [34], where the U orbitals have generally smaller contributions in the bonding molecular orbitals than those of the $C_2$ atoms. There are little differences in the geometrical parameters obtained at PBE0/3-21G and PBE0/6-31G level; however, the charge value of U (-0.208) in linear CUC obtained from PBE0/6-31G is unreasonable, on account of the strong basis set dependence of Mulliken population analyses [49]. The calculated NBO (natural bond orbital) charge population is also unreliable here and the bond distances calculated from the recent developed DFT like M06L functional are overestimated.

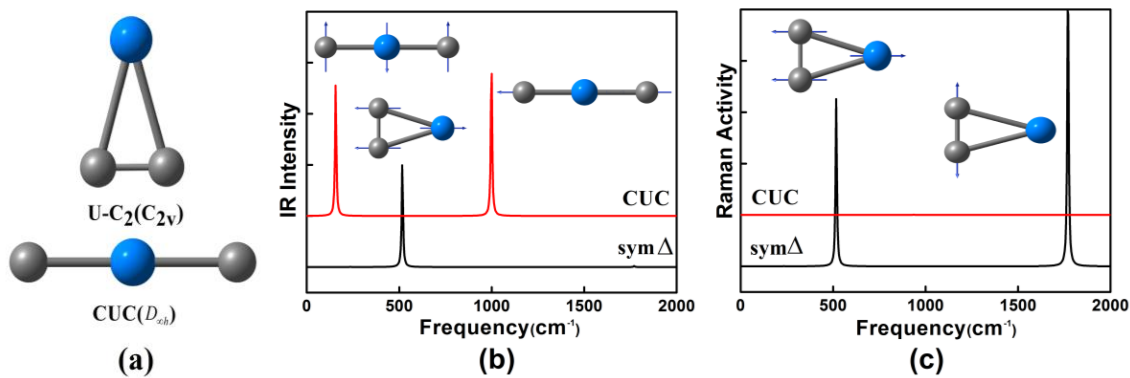

**Figure S1.** Geometries (a) and infrared (b) and Raman (c) vibrational spectra of linear and symmetric triangular $UC_2$ at PBE0/3-21G level.

**Table S5.** Total energies (*E*) of different spin multiplicities (M) for the adsorption systems, and the relative energies (Δ*E*) of the ground state for each system. The results with ZPE corrections are also included.

| System | M | *E* (a.u.) | Δ*E* (eV) | *E*+ZPE (a.u.) | Δ*E*+ZPE (eV) |
|---|---|---|---|---|---|
| I | 3 | -4201.406875 | | | |
| | 5 | **-4201.407926** | **0** | **-4200.532733** | **0** |
| | 7 | -4201.345281 | | | |
| II | 1 | -4201.258331 | | | |
| | 3 | **-4201.272146** | **3.69** | **-4200.393421** | **3.79** |
| | 5 | -4201.250128 | | | |
| III | 1 | -4201.273601 | | | |
| | 3 | **-4201.280027** | **3.48** | **-4200.400215** | **3.61** |
| | 5 | -4201.252229 | | | |
| IV | 1 | -4201.229246 | | | |
| | 3 | **-4201.260222** | **4.02** | **-4200.380859** | **4.13** |
| | 5 | -4201.232882 | | | |

As listed in Table S5, the differences between relative energies with or without ZPE corrections are about 0.1 eV which does not change the order. Moreover, the single-point energy calculation of $E_{relaxed}^{UC_2}$ / $E_{relaxed}^{graphene}$ cannot have the ZPE corrections.

## Part 2. A study of UC$_2$/graphene system at PBE0/6-31G level for comparison

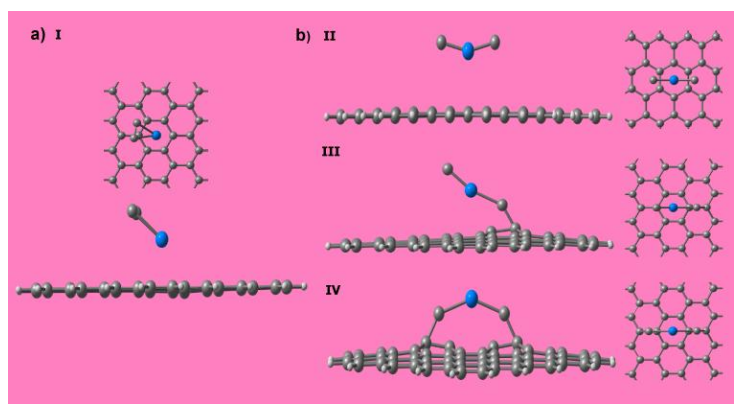

**Figure S2.** Optimized geometries for the graphene-supported UC$_2$ complexes determined at PBE0/6-31G level (top and side views). a) UC$_2$ (C$_{2v}$-symmetric)/graphene, denoted as Structure I. b) UC$_2$ (linear)/graphene, denoted as Structures II, III and IV.

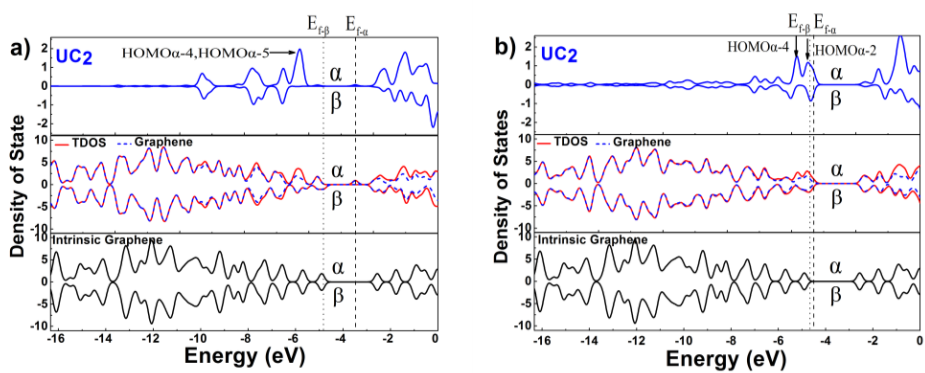

**Figure S3.** DOS of Structure I (a) and Structure IV (b) at PBE0/6-31G level. The E$_{f-\alpha}$ and E$_{f-\beta}$ indicate the Fermi level of α and β electrons of each adsorption system, respectively.

**Table S6.** Relative ($\Delta E$), adsorption ($E_{ads}$), interaction ($E_{int}$), and deformation ($E_{def}$) energies (eV), as well as adsorption distance ($D$, Å) and graphene surface deformation ($\Delta Z$, Å) for the ground state of each adsorption structure at PBE0/6-31G level. All the definitions are same as those in Table 1.

| System | Multiplicity | $\Delta E$ | $E_{ads}$ | $E_{int}$ | $E_{def}^{UC_2}$ | $E_{def}^{graphene}$ | $D$ | $\Delta Z$ |
|---|---|---|---|---|---|---|---|---|
| I | 5 | 0 | 0.92 | 0.99 | 0.01 | 0.07 | 2.46 | 0.14 |
| II | 3 | 3.48 | 0.42 | 0.54 | 0.11 | 0.02 | 2.81 | -0.01 |
| III | 3 | 3.57 | 0.32 | 3.04 | 1.45 | 1.27 | 2.55 | -0.47 |
| IV | 3 | 4.38 | -0.48 | 4.89 | 2.71 | 2.66 | 2.33 | -0.65 |

For a further validation of the approach we used in this work, we recalculated the systems of UC$_2$/graphene at PBE0/6-31G level. As shown in the results of Part 2, although some deviations of the structures and $\Delta Z$ occur (the graphene surfaces in Structures I and II almost locate in a plane, however, bulge more obviously in Structures III and IV), the ground state of each system and information of electronic configuration and molecular orbitals are same as those obtained by 3-21G. Moreover, there are little changes in $\Delta E$. The values of $E_{ads}$ and $E_{int}$ are all small, especially for Structures I and II. The $E_{ads}$ and $E_{int}$ deviate from those in the case showing chemisorption behaviour, as observed in the electronic structures and molecular orbitals of the adsorption systems, confirming the advantage of 3-21G.

## Part 3. Isosurfaces of charge density deformation of Structures I and IV

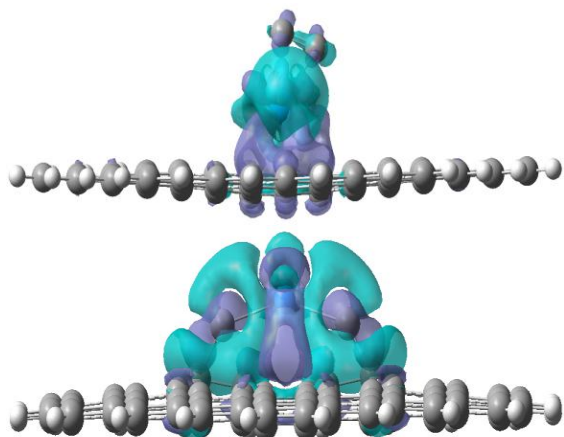

**Figure S4.** Isosurfaces of charge density deformation for Structures I and IV. The blue represents where the electrons are coming from, and the purple shows where the electrons are going to; isosurface value: $\pm 0.002$.

## Part 4. Electronic structure analysis of Structures II and III

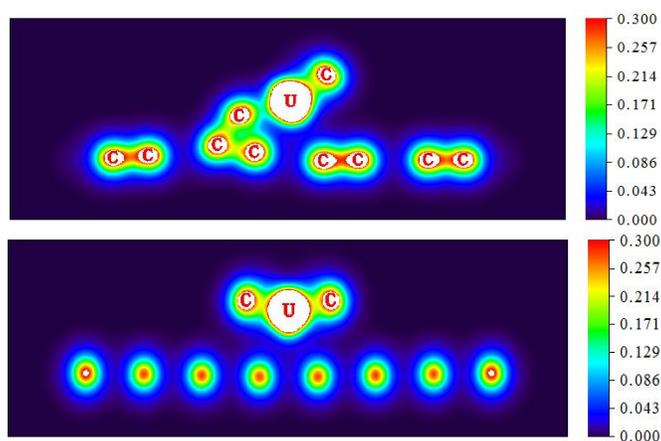

**Figure S5.** Color-filled maps of electron density of Structures II and III (side views).

## Part 5. Molecular orbital diagrams at the adsorption region of Structures I, II and IV.

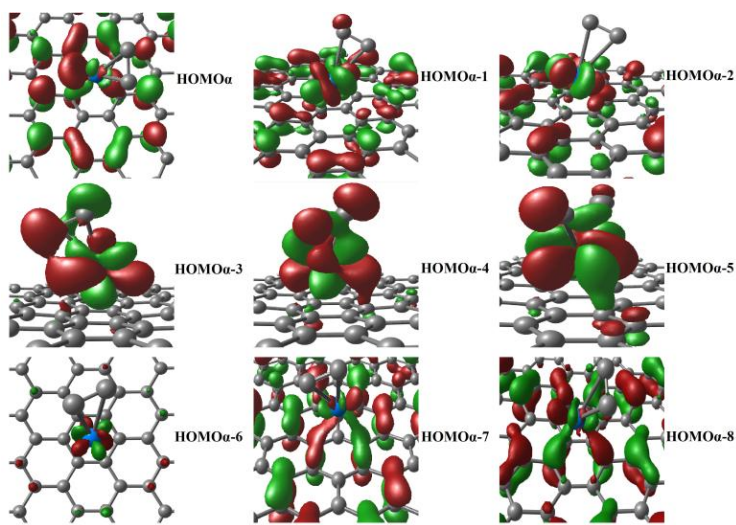

**Figure S6.** Molecular orbital diagrams (isodensity=0.02) at the adsorption region of Structure I.

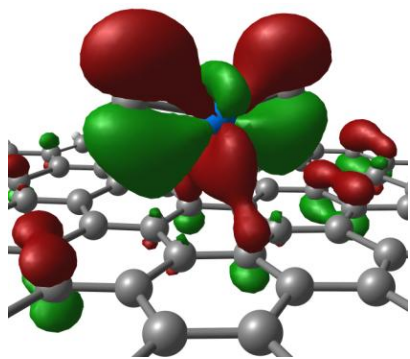

**Figure S7.** Molecular orbital diagram (HOMOα-4, isodensity=0.02) with overlap between the CUC and graphene in Structure II.

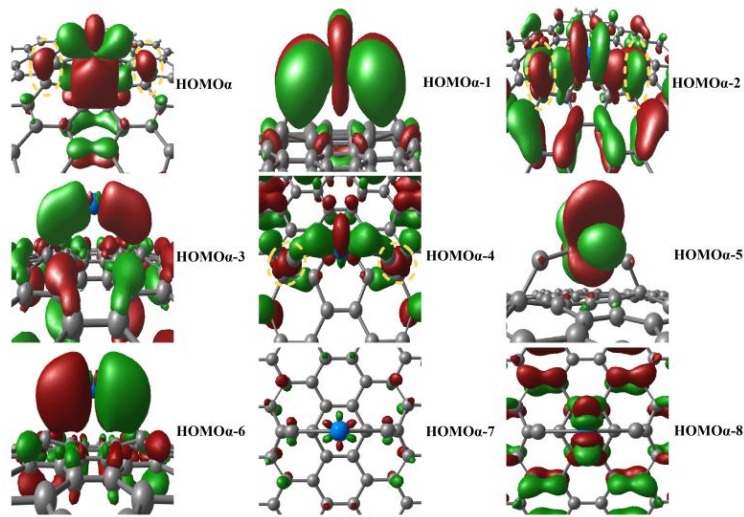

**Figure S8.** Molecular orbital diagrams (isodensity=0.02) at the adsorption region of Structure IV. Circled with yellow dotted lines are double sp$^3$-hybridization sites.

## Part 6. OPDOS between the UC$_2$ and graphene of Structures I (a) and II (b)

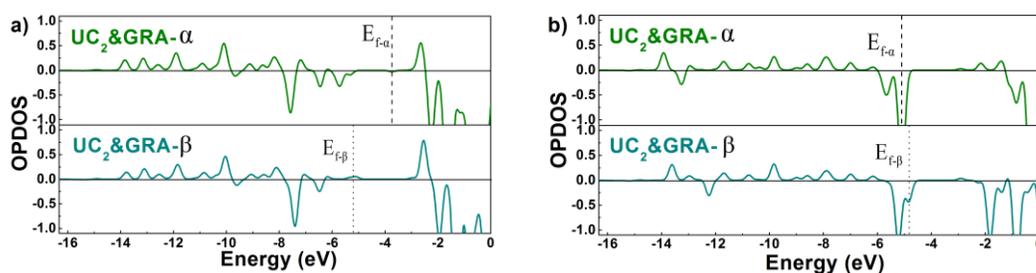

**Figure S9.** OPDOS between the UC$_2$ and graphene of Structures I (a) and II (b). The $E_{f-\alpha}$ and $E_{f-\beta}$ indicate the Fermi levels of α and β electrons, respectively.

The overlap population DOS (OPDOS) demonstrates the relative contribution of the molecular orbitals to the interaction in terms of Mulliken bond order between two orbitals, atoms or groups. The OPDOS shows the bonding, antibonding and nonbonding nature of the interaction of the two fragments, UC$_2$ and graphene, represented by a positive value, a negative value and a zero value, respectively. Although the OPDOS may not offer an accurate quantitative result, we can obtain a qualitative understanding by comparing the bonding contribution area between Structures I and II. As seen in Figure S6, the bonding contribution area at the occupied state in Structure I is greater than that in Structure II. It can explain the reason why the interaction energy of Structure I is higher.

## Part 7. Vibrational modes reflecting the interaction between the UC$_2$ and graphene

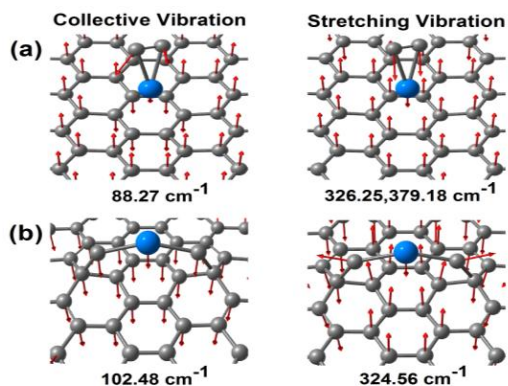

**Figure S10.** The collective and relative stretching vibrational modes on IR spectrum due to interaction between the U and the hexagonal ring of graphene in Structures I (a) and IV (b).

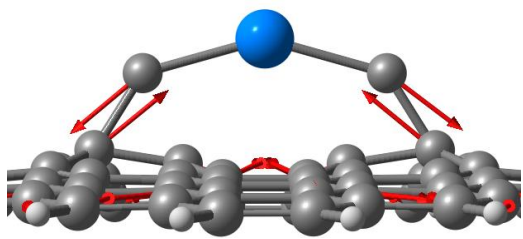

**Figure S11.** The stretching vibration modes between the C at both sides of the UC$_2$ and the C of graphene in Structure IV (986.40 cm$^{-1}$).

## Part 8. Full reference of Ref. 52

Ref. [52]. Gaussian 09, Revision D.01, Frisch, M. J.; Trucks, G. W.; Schlegel, H. B.; Scuseria, G. E.; Robb, M. A.; Cheeseman, J. R.; Scalmani, G.; Barone, V.; Mennucci, B.; Petersson, G. A.; Nakatsuji, H.; Caricato, M.; Li, X.; Hratchian, H. P.; Izmaylov, A. F.; Bloino, J.; Zheng, G.; Sonnenberg, J. L.; Hada, M.; Ehara, M.; Toyota, K.; Fukuda, R.; Hasegawa, J.; Ishida, M.; Nakajima, T.; Honda, Y.; Kitao, O.; Nakai, H.; Vreven, T.; Montgomery, J. A.; Peralta, J. E.; Ogliaro, F.; Bearpark, M.; Heyd, J. J.; Brothers, E.; Kudin, K. N.; Staroverov, V. N.; Keith, T.; Kobayashi, R.; Normand, J.; Raghavachari, K.; Rendell, A.; Burant, J. C.; Iyengar, S. S.; Tomasi, J.; Cossi, M.; Rega, N.; Millam, J. M.; Klene, M.; Knox, J. E.; Cross, J. B.; Bakken, V.; Adamo, C.; Jaramillo, J.; Gomperts, R.; Stratmann, R. E.; Yazyev, O.; Austin, A. J.; Cammi, R.; Pomelli, C.; Ochterski, J. W.; Martin, R. L.; Morokuma, K.; Zakrzewski, V. G.; Voth, G. A.; Salvador, P.; Dannenberg, J. J.; Dapprich, S.; Daniels, A. D.; Farkas, O.; Foresman, J. B.; Ortiz, J. V.; Cioslowski, J.; Fox, D. J, Gaussian, Inc., Wallingford CT, 2013.

## Part 9. References for Supporting Information

[S1]   Moritz, A.; Dolg, M. *Theor. Chem. Acc*, **2008**, *121*, 297-306.